\documentclass{aa}  

\usepackage{graphicx}
\usepackage{txfonts}
\usepackage{natbib}
\bibpunct{(}{)}{;}{a}{}{,}

\begin{document}

\title{A new list of thorium and argon spectral lines in the visible\thanks{Based on measurements made with the HARPS instrument on the ESO 3.6m-telescope at La Silla Observatory under program ID 60.A-9036(A).} \fnmsep \thanks{Table~1 is only available in electronic form at the CDS via anonymous ftp to cdsarc.u-strasbg.fr (130.79.128.5) or via http://cdsweb.u-strasbg.fr/cgi-bin/qcat?J/A+A/}}


\author{C. Lovis
\and F. Pepe
}

\offprints{C. Lovis\\
\email{christophe.lovis@obs.unige.ch}}

\institute{Geneva Observatory, University of Geneva, 51 ch. des Maillettes, 1290 Sauverny, Switzerland
}

\date{Received 6 February 2007 / Accepted 13 March 2007}

\abstract
{}
{We present a new list of thorium and argon emission lines in the visible obtained by analyzing high-resolution ($R=110,000$) spectra of a ThAr hollow cathode lamp. The aim of this new line list is to allow significant improvements in the quality of wavelength calibration for medium- to high-resolution astronomical spectrographs.}
{We use a series of ThAr lamp exposures obtained with the HARPS instrument (High Accuracy Radial-velocity Planet Searcher) to detect previously unknown lines, perform a systematic search for blended lines and correct individual wavelengths by determining the systematic offset of each line relative to the average wavelength solution.}
{We give updated wavelengths for more than 8400 lines over the spectral range 3785--6915~\AA. The typical internal uncertainty on the line positions is estimated to be $\sim$10~m~s$^{-1}$ (3.3 parts in 10$^8$ or 0.18~m\AA), which is a factor of 2--10 better than the widely used Los Alamos Atlas of the Thorium Spectrum (Palmer \& Engleman 1983). The absolute accuracy of the global wavelength scale is the same as in the Los Alamos Atlas. Using this new line list on HARPS ThAr spectra, we are able to obtain a global wavelength calibration which is precise at the 20~cm~s$^{-1}$ level (6.7 parts in 10$^{10}$ or 0.0037~m\AA).}
{Several research fields in astronomy requiring high-precision wavelength calibration in the visible (e.g. radial velocity planet searches, variability of fundamental constants) should benefit from using the new line list.}


\maketitle

\section{Introduction}

The wavelength calibration of astronomical spectrographs in the visible is usually performed using arc lamp spectra such as those produced in hollow cathode lamps. Thorium is today the most often used element due to its numerous spectral features over the whole visible and near-infrared domains. Other advantages of thorium include its mono-isotopic nature and the absence of hyperfine structure, which lead to narrow, highly symmetric line profiles. For all these reasons, ThNe and ThAr hollow cathode lamps have become the standard for wavelength calibration of astronomical spectrographs.

The widely used reference for accurate Th wavelengths is the Atlas of the Thorium Spectrum obtained with the McMath-Pierce Fourier Transform Spectrometer (FTS) of the National Solar Observatory at Kitt Peak \citep[][hereafter PE83]{palmer83}. The FTS scans from which this atlas was derived have a spectral resolution of about 600,000. The atlas gives the position of about 11,500 lines between 3000 and 11,000~\AA. The quoted wavenumber accuracy ranges from 0.001~cm$^{-1}$ for the strongest lines to 0.005~cm$^{-1}$ for the faintest ones. In velocity units at 5500~\AA, this corresponds to 16--82~m~s$^{-1}$ (5.3--27 parts in 10$^8$ or 0.29--1.50~m\AA).

While the wavelength accuracy of the PE83 atlas will be high enough for most astronomical applications, it starts to be a limiting factor when aiming at very high-precision radial velocity (RV) measurements. Over the past decade, the need for such measurements has considerably increased following the discoveries of the first extrasolar planets using Doppler velocimetry. The improvements of this technique over the years have led to the detection of planets having the mass of Neptune and below. This requires a Doppler precision of 1~m~s$^{-1}$ or less \citep[e.g.][]{lovis06}. To carry out such observations successfully, all aspects influencing the RV measurement, from the properties of the source to the instrumental design and the data reduction, have to be carefully taken into account. As far as we are concerned, we have been particularly involved in the development and operation of the HARPS instrument, a cross-dispersed, fiber-fed echelle spectrograph mounted on the European Southern Observatory 3.6m-telescope at La Silla, Chile \citep[][]{mayor03}. HARPS is operated under vacuum in an isolated, strictly temperature-controlled environment and is therefore extremely stable and well suited to high-precision spectroscopic work. We have investigated many aspects regarding instrument calibration and data reduction, and identified several domains where improvements could be made. One of these is the wavelength calibration process.

Two main techniques are commonly used today to achieve high-precision RV measurements: the so-called iodine cell \citep[][]{butler96} and simultaneous ThAr \citep[][]{baranne96} techniques. While the former uses a superimposed iodine spectrum as absolute wavelength reference, the latter relies on ThAr calibration spectra (this is the one we use on HARPS). In this case, the stability and precision of the wavelength solutions derived from ThAr spectra is a crucial aspect of the calibration process. It turns out that the precision of the Th wavelengths as given in PE83 is no longer adequate for the wavelength calibration of high-resolution spectrographs if a RV precision of 1~m~s$^{-1}$ is to be reached. Indeed, errors in the PE83 atlas, which are due to random noise in the FTS scans, translate into systematic offsets of the lines with respect to the average wavelength solution in the spectral chunk to be calibrated. The residuals around the wavelength solution are no longer dominated by photon shot noise, but by these systematic offsets which can be one order of magnitude larger than random noise, as for example in the case of HARPS. As a result, wavelength solutions will not be well constrained and will show unstable behaviour against adding or removing Th lines in the fitting process.

The limited precision of the PE83 wavelengths led us to the idea of creating a new atlas of Th and Ar lines based on HARPS ThAr spectra. Contrary to Fourier transform spectrometers, grating instruments have complicated dispersion relations and their overall wavelength scale cannot be determined from the measurement of a single well-known transition. However, they are usually much more sensitive, which should make it possible to use such instruments to reduce the random noise on the PE83 wavelengths while keeping the absolute wavelength scale given by the FTS. HARPS spectra are indeed of sufficient quality to permit significant improvements in the relative precision on line positions. Moreover, these spectra will allow us to identify thousands of unknown (mostly faint) lines, which are not present in the PE83 atlas due to its lower sensitivity. Finally, this would give us the opportunity to perform a careful treatment of blended lines, which represent a major problem when aiming at high-precision wavelength calibration.

In this paper we describe the new line list we obtained using HARPS ThAr spectra. The different steps of the process are detailed in Sect.~2, and the results presented in Sect.~3. A discussion about the potential of the new atlas, its advantages and limitations, follows in Sect.~4.

\section{Experimental method}

\subsection{Instrumental setup}

The light source for the Th spectrum is a standard ThAr hollow cathode lamp\footnote{The lamp was manufactured by Cathodeon/Heraeus (type 3UAX/Th, serial number B13773).} operated at a current of 9.0~mA. The lamp window, made of UV glass, has a diameter of 37~mm. The pressure of the fill gas is 800~Pa. In such lamps, a discharge is established between the anode and the Th cathode through the carrier gas (Ar in our case). Collisions between the accelerated ions and the cathode sputter off Th atoms from the cathode. Collisions in the plasma in turn excite Th atoms, which emit light by decaying to lower states. As a result, a rich spectrum of Th and Ar emission lines is produced.

The beam from the ThAr lamp is injected into an optical fiber carrying the light to the spectrograph, which is located inside a vacuum vessel in an isolated room below the telescope. Light is dispersed on a large R4 echelle grating and the echelle orders are cross-dispersed by a grism. The entrance slit (i.e. the fiber output) is then re-imaged onto a 4k4k CCD mosaic, made of two 2k4k EEV CCDs with a pixel size of 15 $\mu$m. The spectral format consists of 72 echelle orders covering the whole visible spectral range from 3785 to 6915~\AA, except the interval 5306--5339~\AA\ which falls within the gap between both CCDs. The spectral resolution is $R=110,000$ and the average dispersion 0.015~\AA\,pixel$^{-1}$ (820~m~s$^{-1}$~pixel$^{-1}$ in velocity units). This leads to a sampling of $\sim$3.3~pixels per resolution element.

\subsection{Characteristics of HARPS ThAr spectra}

In the following we will be using a series of ThAr lamp exposures obtained in December 2004 as part of the standard calibration plan. The exposure time was 40~s for each frame. ThAr spectra have been extracted and flat-fielded using the standard echelle reduction procedures implemented in the HARPS data reduction software. The basic data products we are starting from are the extracted two-dimensional (pixel vs. order) spectra. These contain thousands of Th lines, which are unresolved at the HARPS resolution and intrinsically highly symmetric. Moreover, the instrumental profile itself is very close to Gaussian. In the following we therefore use simple Gaussian functions to fit the position of the lines on the extracted spectra.

Line intensities extend over the whole dynamic range of the CCD. For the strongest unsaturated lines, the photon noise on the line position is about 0.002~pixel, corresponding to 1.6~m~s$^{-1}$ on average (5.3 parts in 10$^9$ or 0.029~m\AA). On the other hand, the position of the faintest detectable lines can be measured with a precision of about 0.2~pixel (160~m~s$^{-1}$). However, in the case of very strong lines, photon noise is actually not the limiting factor. We estimate that CCD inhomogeneities, such as pixel size variations (up to a few percent), inhomogeneous intra-pixel sensitivity or CCD block stitching errors, will limit the precision at which the true position of a line can be measured to about 0.01~pixel, corresponding to 8~m~s$^{-1}$ (2.7 parts in 10$^8$ or 0.15~m\AA).

\begin{figure}
\resizebox{\hsize}{!}{\includegraphics[angle=270]{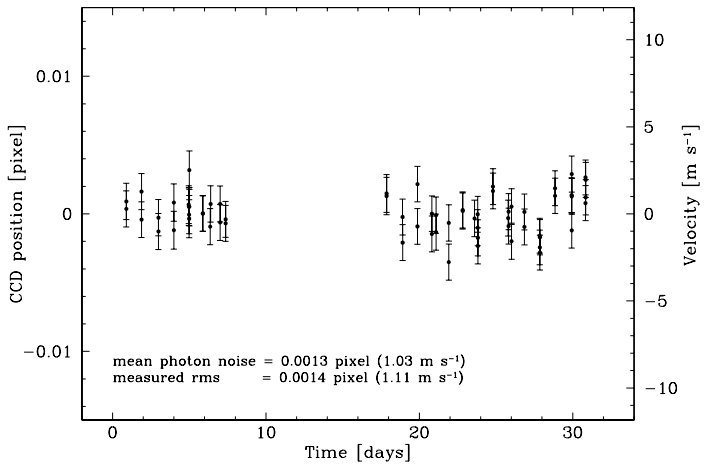}}
\caption{HARPS instrumental stability and precision on the measurement of line positions. The CCD position of a strong Th line is measured on 64 different ThAr lamp exposures taken over a period of one month. The measured scatter shows that the spectrograph did not drift during this period within the error bar of less than 0.002~pixel.}
\label{FigStability}
\end{figure}

To minimize photon noise, the most straightforward option would be to combine many ThAr spectra and measure averaged line positions. However, this is only feasible if the instrument remains sufficiently stable over the time duration of the ThAr sequence. This condition is usually not fulfilled under normal terrestrial conditions because variations in atmospheric pressure, temperature changes and mechanical flexures induce instrumental drifts of tens to hundreds m~s$^{-1}$ over timescales of minutes to hours. In the case of HARPS, instrumental stability is such that the spectrum does not drift by more than 1~m~s$^{-1}$ over timescales of days to months (see Fig.~\ref{FigStability}). This allowed us to combine a series of 64 ThAr spectra taken over one month to measure line positions, thereby considerably reducing photon noise. For the vast majority of the lines, the measurement uncertainty could actually be reduced to about 10~m~s$^{-1}$ (3.3 parts in 10$^8$ or 0.18~m\AA). A quick comparison with the numbers quoted for the PE83 atlas in Sect.~1 shows that line positions can be determined 2--10 times more precisely using HARPS ThAr spectra. The use of the PE83 atlas leads to a dispersion of the residuals around wavelength solutions of 50--70~m~s$^{-1}$ due to the errors in the input wavelengths. The creation of a new line list would allow us to bring this number down to about 10~m~s$^{-1}$. This is the main motivation for the new atlas.

\subsection{Line-fitting process}

At the beginning of the process, we developed an algorithm to systematically search for lines in the ThAr spectra in order to include several thousands of lines which are not listed in the PE83 atlas because it is less sensitive than our HARPS spectra (except in the bluest part). The use of a much larger number of lines, even if most of them are quite faint, has a stabilizing effect on the wavelength solutions, especially in the red part of the spectrum where strong Th lines are scarce. Thanks to the newly detected lines we were able to increase the total number of lines by~$\sim$50\%.

\begin{figure}
\resizebox{\hsize}{!}{\includegraphics{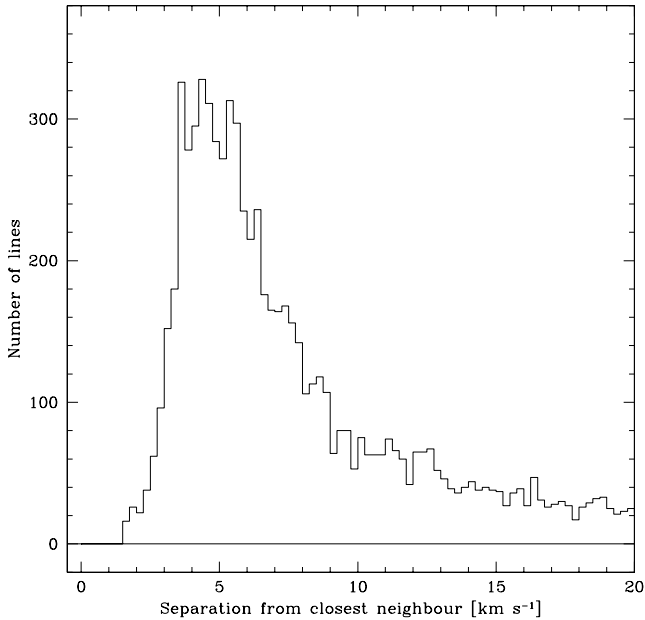}}
\caption{Closest-neighbour separation between lines in the final list.}
\label{FigSeparation}
\end{figure}

This has a major impact on the occurrence of blends. ThAr spectra are actually so rich that truly isolated lines are very rare, even at a resolution of 110,000. Blends represent a major problem for high-precision wavelength calibration because they induce shifts in the centroids of the components which are often much larger than photon noise. Fig.~\ref{FigSeparation} shows a histogram of the closest-neighbour separation between lines. The histogram peaks at $\sim$4.5~km~s$^{-1}$, which corresponds to $\sim$1.5~times the spectral resolution. This obviously means that there remain many undetected blends within the spectral resolution. However, we automatically rejected lines that appear single in our spectra, but actually contain other components resolved by PE83 (at $R=600,000$). In this way we eliminated most of the dangerous blends. The effects of unidentified, faint blending components within the HARPS resolution are less severe since they will mostly induce shifts comparable to, or smaller than, photon noise. To deal with blended lines partially resolved by HARPS, we developed an algorithm to simultaneously fit neighbouring lines with multiple Gaussians. The algorithm determines which lines cannot be fitted independently and groups them with their closest neighbours to make a multiple fit. This allowed us to keep in our list blended lines whose components are only $\sim$1~resolution element apart (2.7~km~s$^{-1}$), provided the components are of similar intensity.

While the occurrence of too close blends was the main cause for rejecting lines, about 100--200 strongly saturated lines and those in their neighbourhood had to be removed as well because they were impossible to fit properly. At the other extremity of the intensity scale, some PE83 lines were too faint to be fitted, especially in the bluest part of our spectra.

After completion of the list of lines that could be properly fitted, we ran the line-fitting algorithm on each of the 64 ThAr spectra of our series. We then computed for each line its mean position, rejecting outlying measurements using sigma-clipping. Lines which showed an unusual number of rejected measurements were removed from the list for being unstable to the fit. Reasons for unstable behaviour include blended components with large intensity differences, faint lines at the detection limit, pollution by neighbouring strongly saturated lines, possible intrinsic instability of a few contaminant lines, etc. The resulting list is our starting point for the wavelength calibration process.

\subsection{Determination of updated wavelengths}

The basic principle for the determination of updated wavelengths is extremely simple: a wavelength solution is fitted through the measured line positions (in pixels) using their PE83 wavelengths. The residuals to the fit will not be dominated by random noise from HARPS spectra, but by systematics from the PE83 atlas. Corrections to the PE83 wavelengths are then simply given by the residuals to the wavelength solution. Obviously, wavelength solutions themselves have to be sufficiently well constrained and match the true optical dispersion relation as closely as possible. However, the systematics we want to correct were originally nothing but random noise in the FTS scans from which the PE83 atlas was derived. They are therefore randomly distributed as a function of wavelength and do not have any large-scale impact. The best wavelength solution will then still be the one which minimizes $\chi^2$, i.e. the sum of the (properly weighted) squared residuals around the fit.

We emphasize that this method will yield updated wavelengths which are {\it internally} more precise than in PE83, but the global wavelength scale itself will be the same as in PE83, since we made no attempt to improve the {\it absolute} wavelength accuracy. However, this has no impact on high-precision RV measurements for example, since these are essentially differential measurements.

Our ThAr spectra are split into 72 spectral orders that overlap each other to some extent. This means that a significant fraction of Th lines are detected twice (on two consecutive orders). If we now fit an independent wavelength solution on each order, we will find two different systematics for a given line in the overlap region. This is particularly unsatisfactory because it is precisely towards order edges that wavelength solutions are most poorly constrained. We therefore chose to couple the wavelength solutions of all orders by imposing one and the same systematic for all lines appearing twice. In practice, we first had to determine which lines are detected twice by assigning a preliminary wavelength to all of them. This was done by fitting independent wavelength solutions on all orders. We then classified lines as being unique or duplicate. Concerning wavelength solutions themselves, we found that third-order polynomials $\lambda(x) = \sum_{k=0}^{3}{a_k x^k}$ give an adequate analytical representation in the sense that the residuals around the fit are randomly distributed as a function of wavelength (no large-scale trends). Moreover, supplementary degrees of freedom would often increase the reduced~$\chi^2$ and are therefore not desirable.

The coupling between all orders generates a large linear least-squares problem where lines contribute different types of equations depending on their classification:

\begin{itemize}

\item Duplicate PE83 lines: two equations of type
\begin{equation}
\label{type1}
\sum{a_k x_i^k} - \Delta\lambda_i = \lambda_{i,\mathrm{PE83}}
\end{equation}

\item Unique PE83 lines: one equation of type
\begin{equation}
\label{type2}
\sum{a_k x_i^k} = \lambda_{i,\mathrm{PE83}}
\end{equation}

\item Duplicate unidentified or Ar lines: two equations of type
\begin{equation}
\label{type3}
\sum{a_k x_i^k} - \lambda_i = 0
\end{equation}

\item Unique unidentified or Ar lines: no equation

\end{itemize}

In the above equations, the $a_k$ are the unknown polynomial coefficients of the wavelength solution for the corresponding order, $x_i$ is the position of line~$i$ in pixels, $\Delta\lambda_i$ is the unknown systematic offset of line~$i$, $\lambda_{i,\mathrm{PE83}}$ is the PE83 wavelength of line~$i$ and $\lambda_i$ is the unknown wavelength of line~$i$ (for unidentified or Ar lines). We consider Ar lines as having unknown wavelengths because they do not appear in PE83 (which used a ThNe source) and because they are sensitive to physical conditions in the lamp (see Sect.~3). The individual systematics $\Delta\lambda_i$ are unknowns to be determined, but they must have a global property that we can use to introduce more constraints on them in the least-squares minimization: they should be normally distributed around zero. We can therefore add a further type of equation (one for each $\Delta\lambda_i$):

\begin{equation}
\label{type4}
\hspace{0.5cm} \Delta\lambda_i = 0
\end{equation}

This has to be understood in a least-squares sense: these equations will ensure that, in the global solution, the properly weighted (see below) $\Delta\lambda_i$ will have a mean value close to zero and a variance close to one.

We now need to assign the correct weight to each equation so that we can solve our weighted linear least-squares problem. Uncertainties for each equation will be computed differently depending on equation type. For types~\ref{type1} and~\ref{type3}, uncertainties are simply given by random noise from HARPS spectra, derived from the combination of the 64 individual spectra. They are computed as the quadratic sum of photon noise and detector-related noise (0.01~pixel). For most lines these error bars could be reduced below 20~m~s$^{-1}$. For equation type~\ref{type2}, uncertainties are given by the quadratic sum of HARPS random noise and the PE83 wavelength uncertainty for the line under consideration (see below). Finally, equations of type~\ref{type4} will be weighted with the PE83 uncertainty alone.

\begin{figure}
\resizebox{\hsize}{!}{\includegraphics{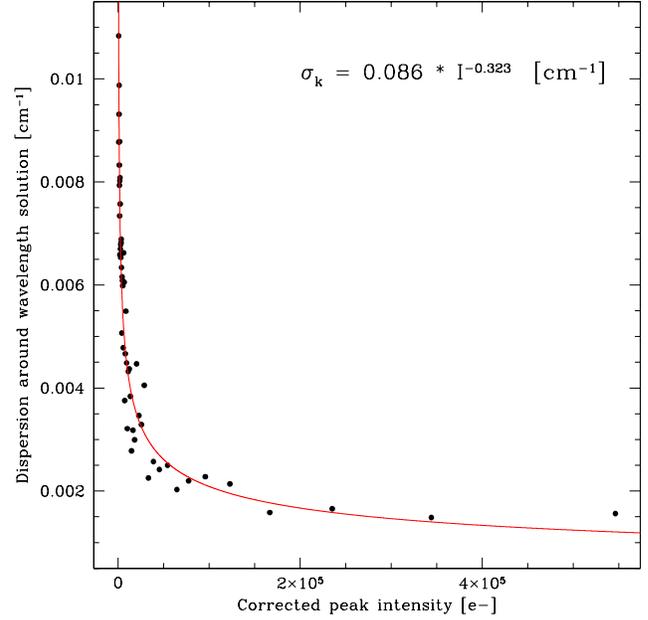}}
\caption{Relation between line intensities and the dispersion of residuals around wavelength solutions when using the PE83 atlas. This gives the errors on the PE83 wavelengths as a function of the measured line intensity on HARPS spectra. In velocity units at 5500~\AA, errors range from 20~m~s$^{-1}$ to $\sim$160~m~s$^{-1}$ (6.7 to 53 parts in 10$^8$ or 0.37 to 2.9~m\AA).}
\label{FigPE83error}
\end{figure}

To determine the PE83 wavelength uncertainty for each line, we first computed the residuals of the PE83 lines relative to the wavelength solutions independently obtained on each spectral order. We then plotted these residuals as a function of the line peak intensities, which were previously corrected for the wavelength-dependent HARPS instrumental efficiency. The resulting plot is shown in Fig.~\ref{FigPE83error}. Residuals are given in~cm$^{-1}$ because the PE83 uncertainties expressed in these units are expected to be almost independent on wavelength (see PE83). There is a clear relation between line intensities and residuals, which actually just reflects the photon noise on the FTS scans from which the PE83 atlas was derived. We fitted a power law to these data to obtain an analytical expression relating the PE83 uncertainty to the measured line intensity on HARPS spectra. This relation is in very good agreement with the error bars quoted in PE83 (see Sect.~1).

The weighted linear least-squares problem can now be solved. We have in total 7648 equations and 2082 unknowns. The $\chi^2$-minimization yields best-fit values for the coefficients of the wavelength solutions $a_k$ and the systematics $\Delta\lambda_i$, from which updated wavelengths can be obtained for all PE83, unidentified and Ar lines. The reduced~$\chi^2$ is equal to~1.57. Several factors can explain its somewhat high value, among which a slight underestimation of the detector-related noise on HARPS spectra, a non-perfect analytical model to describe wavelength solutions, and the presence of undetected blends that introduce small shifts in line centroids. We checked that varying slightly the error bars has a negligible impact on the final results. Consequently we preferred to keep our best error estimates rather than artificially increase them.

\section{The new line list}

\begin{table*}
\caption{The first 20~lines of the new atlas. See text for a description of the different columns.}
\label{TableNewAtlas}
\centering
\begin{tabular}{c c c c c c}
\hline\hline
New vacuum wavelength & Uncertainty & Wavenumber & Intensity & Previous wavelength & Identification \\
$[$\AA$]$ & $[$\AA$]$ & [cm$^{-1}$] & [e-] & $[$\AA$]$ & \\
\hline
3785.650454 & 0.000149 & 26415.53974 & 7273   & 3785.649743 & NoID \\
3786.675812 & 0.000129 & 26408.38692 & 11987  & 3786.675314 & ThII \\
3787.458325 & 0.000120 & 26402.93078 & 24284  & 3787.457347 & ArII \\
3788.714137 & 0.000160 & 26394.17923 & 3050   & 3788.714472 & ThI \\
3789.009459 & 0.000170 & 26392.12203 & 2584   & 3789.010941 & ThI \\
3789.435335 & 0.000157 & 26389.15595 & 2707   & 3789.435169 & ThII \\
3790.185751 & 0.000527 & 26383.93118 & 6700   & --          & ? \\
3790.243720 & 0.000114 & 26383.52765 & 126158 & 3790.243943 & ThI \\
3791.160081 & 0.000128 & 26377.15049 & 4572   & 3791.158240 & NoID \\
3791.343954 & 0.000146 & 26375.87125 & 2491   & 3791.344838 & ThI \\
3791.431959 & 0.000119 & 26375.25903 & 16543  & 3791.432308 & NoID \\
3791.496302 & 0.001561 & 26374.81143 & 1711   & --          & ? \\
3791.871484 & 0.000113 & 26372.20181 & 68571  & 3791.871327 & ThI \\
3792.296894 & 0.000140 & 26369.24344 & 2497   & 3792.297101 & ThI \\
3792.373700 & 0.000149 & 26368.70939 & 2076   & 3792.373871 & ThII \\
3792.594893 & 0.000126 & 26367.17151 & 4127   & 3792.594391 & ThI \\
3793.807442 & 0.000114 & 26358.74422 & 13282  & 3793.806899 & ThI \\
3794.893018 & 0.000118 & 26351.20398 & 4508   & 3794.893246 & ThI \\
3795.228217 & 0.000125 & 26348.87661 & 2699   & 3795.228593 & ThII \\
3795.775429 & 0.000112 & 26345.07807 & 11036  & 3795.775842 & ThI \\
\hline
\end{tabular}
\end{table*}

The new list gives updated wavelengths for 8442~lines between 3785 and 6915~\AA. The data, available in electronic form at the CDS, contain 6~columns, as shown in Table~\ref{TableNewAtlas}. The first column gives the new vacuum wavelength of the lines as obtained in this work. The second column contains the estimated internal uncertainties on line positions. The third column gives the wavenumbers corresponding to the new wavelengths. Peak intensities, corrected for instrumental response, are listed in the fourth column. The fifth column gives the previously published wavelengths when available. For Th lines, this is the PE83 wavelength. For Ar lines, wavelengths have been taken from \citet{norlen73} and \citet{whaling95,whaling02}. For unidentified lines, the value was set to '--'. Finally, the last column gives line identifications when available. 'ThI', 'ThII', 'ThIII', 'NoID' and 'Contam' are all identifications from PE83. The first three are self-explanatory. 'NoID' stands for lines that could not be identified as known Th transitions by PE83, but are probably due to Th. 'Contam' stands for identified contaminants. 'ArI' and 'ArII' identifications are from \citet{norlen73} and \citet{whaling95,whaling02}. Previously unidentified lines are designated with~'?'. It should be noted that the line intensities given in the table are only indicative since flux ratios between lines may exhibit variations from one ThAr lamp to the other. In particular, the Ar-to-Th flux ratio is very sensitive to physical conditions in the lamp.

We made some comparisons between several lamps of the same manufacturer to check whether wavelengths themselves might be affected by changing physical conditions in the lamps. Although a detailed analysis is beyond the scope of this paper, two main conclusions can already be drawn. First, Th lines turn out to be individually stable in wavelength at least within the error bars quoted in this paper. Second, Ar lines exhibit wavelength shifts from lamp to lamp amounting to a few tens of~m~s$^{-1}$. This behaviour is probably the result of pressure shifts, to which the Ar atom is known to be quite sensitive \citep[see][]{whaling02}. As a consequence, the Ar wavelengths given in the atlas have to be taken with caution. In general, we strongly recommend not to use Ar lines to perform wavelength calibration if high precision is needed.

\section{Discussion and conclusion}

\subsection{Comparison with previous line lists}

\begin{figure}
\resizebox{\hsize}{!}{\includegraphics{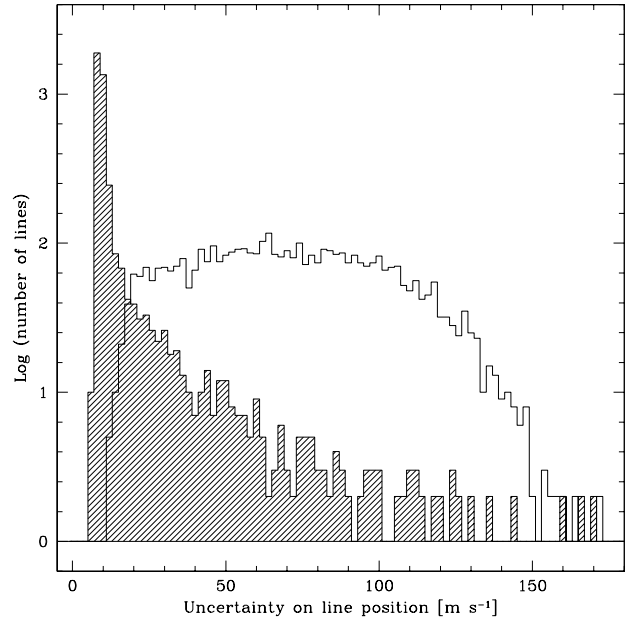}}
\caption{Distribution of uncertainties on line positions in the new atlas (shaded histogram) and the PE83 atlas (plain histogram). Note the logarithmic scale on the vertical axis.}
\label{FigErrorDistributions}
\end{figure}

If we merge the best available Th and Ar line lists in the literature \citep[][]{palmer83,norlen73,whaling95,whaling02}, the total number of lines in the interval 3785--6915~\AA\ amounts to~7321. However, an important fraction of these cannot be used when working at a resolution of 110,000 because of blending of individual components. Actually, the number of usable lines from these lists amounts to about~3900. The new atlas contains $\sim$4000 previously unidentified lines and $\sim$500 known lines which could not be used before because of blends with unknown components. In total, the new list more than doubles the number of lines available for wavelength calibration.

As far as precision is concerned, we have briefly described the error distributions in PE83 and the new atlas in Sect.~2. To illustrate this more completely, Fig.~\ref{FigErrorDistributions} shows both distributions on the same plot. The gain in precision is clearly seen and fully justifies the creation of the new atlas.

\begin{figure}
\resizebox{\hsize}{!}{\includegraphics[angle=270]{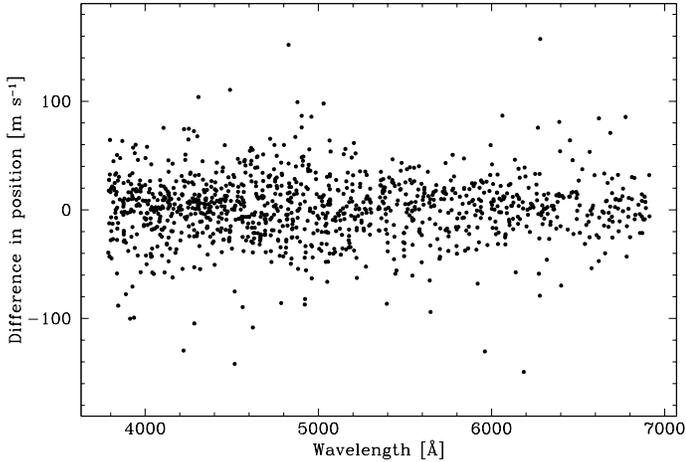}}
\caption{Difference in line positions between the new atlas and the PE83 atlas as a function of wavelength. Only lines with a PE83 error below 50~m~s$^{-1}$ are shown. Both atlases have the same absolute wavelength scale.}
\label{FigDvNewOld}
\end{figure}

We mentioned in Sect.~2 that the new list has, by construction, the same absolute, global wavelength scale as PE83. This is illustrated in Fig.~\ref{FigDvNewOld}, which shows the difference in line positions between both atlases. As expected, no large-scale trend can be seen. The mean offset between both lists amounts to only 0.08~m~s$^{-1}$ (2.7 parts in 10$^{10}$ or 0.0015 m\AA). The scatter itself mainly reflects the errors in PE83.

\subsection{Precision of wavelength calibration}

\begin{figure}
\resizebox{\hsize}{!}{\includegraphics[angle=270]{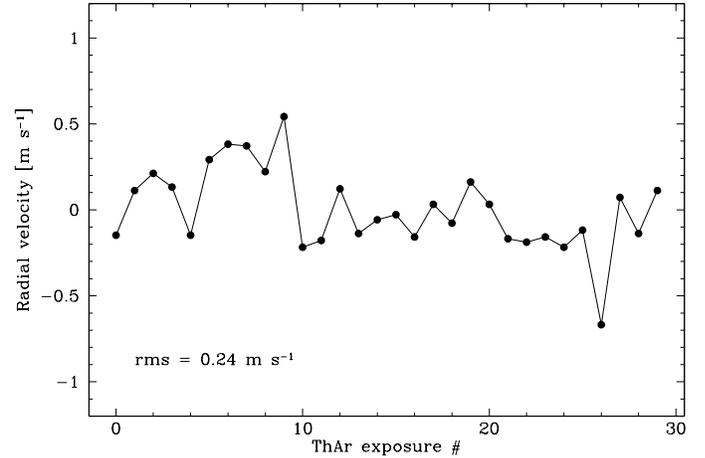}}
\caption{Test of the HARPS wavelength calibration accuracy. The radial velocity of a star is computed from a single stellar exposure using 30 different ThAr exposures as calibration frames. The obtained RV dispersion is a measure of the precision achieved by the wavelength calibration process.}
\label{FigZeroPoint}
\end{figure}

As a final step of our work, we implemented a new wavelength calibration algorithm in the HARPS data reduction pipeline in order to use the new line list. We now routinely obtain residuals around the fit which have a weighted rms dispersion of about~10~m~s$^{-1}$ (3.3 parts in 10$^8$ or 0.18 m\AA), as expected from the combination of photon noise and uncertainties in the input line list. This means that the wavelength solution is precise at a level of a few~m~s$^{-1}$ locally, and at the unprecedented level of 20~cm~s$^{-1}$ (6.7 parts in 10$^{10}$ or 0.0037 m\AA) globally (i.e. for example when computing a radial velocity using the whole spectral range). This can be demonstrated in the following way: we choose one observation of a star and reduce it many times using different wavelength calibrations, i.e. different ThAr exposures. This will yield each time a slightly different radial velocity for the star. The dispersion of these RVs is a measure of the precision we have on the wavelength solutions. Fig.~\ref{FigZeroPoint} shows the results of this test. Using a series of 30~ThAr exposures, we obtain a raw dispersion of 24~cm~s$^{-1}$. Correcting for instrumental drifts at the level of~$\sim$10~cm~s$^{-1}$ occurring during the series, this number reduces to 20~cm~s$^{-1}$.

As a conclusion, we find that the quality of the new line list permits significant improvements in the wavelength calibration process in general. A number of research fields in astronomy requiring high-precision wavelength calibration will certainly benefit from using the new atlas. Apart from radial velocity searches for extrasolar planets, we can also mention the work on the variability of fundamental constants \citep[e.g.][]{murphy03,levshakov05,chand06}, where an accurate wavelength scale is needed to properly measure line shifts possibly caused by varying physical constants over cosmological timescales.

We recommend using the new list for calibrating all spectrographs working at a resolution $R \leq 110,000$. Obviously, the lower the resolution, the more adaptations will be needed to deal with blended lines. For each blend, the expected shift of the main component can be estimated using the positions and intensities of all components given in the input list. This value can then be compared to the random noise on the main component (or to some tolerance value) to decide whether or not the line can be safely used for wavelength calibration. An example of such a procedure illustrating how to select suitable lines at lower resolution is presented in \citet{murphy07a}. At resolutions higher than 110,000, the choice of the input line list depends on what is exactly needed. The new atlas should be sufficient to guarantee a precise wavelength calibration provided the number of detected lines is high enough. If more lines are needed, then it is necessary to include lines from PE83 that were removed from the new atlas because of blends. In this case, great care should be taken in selecting lines to make sure they are not blended with unknown components.

\subsection{Future developments}

Although ThAr lamps have several qualities as wavelength calibrators, they also have a number of drawbacks. At moderate spectral resolution, blends represent a major problem which may prevent high-precision wavelength calibration. Furthermore, the high dynamic range in line intensities makes it sometimes difficult to find a good compromise between, on the one hand, the minimization of strongly saturated lines and, on the other hand, the availability of a large number of lines.

Finally, the finite lifetime of lamps and possible aging effects affecting Th wavelengths at the~$\sim$1~m~s$^{-1}$ level make these lamps unsuitable for future instruments aiming at~cm~s$^{-1}$ precision. This is for example the case for CODEX (COsmic Dynamics EXperiment), a project for a high-resolution spectrograph to be installed on the European Extremely Large Telescope \citep[][]{pasquini06,grazian07}. New calibration systems have to be developed to ensure a wavelength accuracy and stability at the~10$^{-11}$ level. This has to be compared to the accuracy level of a few parts in 10$^8$ achieved by PE83. The needed "ideal" calibrator would produce a dense pattern of equally spaced lines with precisely known and stable wavelengths. Moreover, the lines should all have about the same intensity. All these requirements are obviously challenging, but efforts are presently being made to adapt the laser frequency comb technology to these needs \citep[see][]{murphy07b}. The possibility to couple the laser comb to an atomic clock would permit to reach the required very high stability level. There is little doubt that, once fully adapted, such a calibration system would find applications in a number of different domains.

\begin{acknowledgements}

We would like to thank the Swiss National Science Foundation (FNRS) for its continuous support. We also thank M.~Murphy for providing us with the PE83 and Ar line lists in electronic form. We are grateful to G.~Lo~Curto and A.~Gilliotte for their support in testing the HARPS ThAr lamps.

\end{acknowledgements}

\bibliographystyle{aa}
\bibliography{biblio}

\begin{thebibliography}{15}
\expandafter\ifx\csname natexlab\endcsname\relax\def\natexlab#1{#1}\fi

\bibitem[{{Baranne} {et~al.}(1996){Baranne}, {Queloz}, {Mayor}, {Adrianzyk},
  {Knispel}, {Kohler}, {Lacroix}, {Meunier}, {Rimbaud}, \& {Vin}}]{baranne96}
{Baranne}, A., {Queloz}, D., {Mayor}, M., {et~al.} 1996, \aaps, 119, 373

\bibitem[{{Butler} {et~al.}(1996){Butler}, {Marcy}, {Williams}, {McCarthy},
  {Dosanjh}, \& {Vogt}}]{butler96}
{Butler}, R.~P., {Marcy}, G.~W., {Williams}, E., {et~al.} 1996, \pasp, 108, 500

\bibitem[{{Chand} {et~al.}(2006){Chand}, {Srianand}, {Petitjean}, {Aracil},
  {Quast}, \& {Reimers}}]{chand06}
{Chand}, H., {Srianand}, R., {Petitjean}, P., {et~al.} 2006, \aap, 451, 45

\bibitem[{{Grazian} {et~al.}(2007){Grazian}, {Vanzella}, {Cristiani},
  {Dessauges-Zavadsky}, {D'Odorico}, M.G., \& {et al.}}]{grazian07}
{Grazian}, A., {Vanzella}, E., {Cristiani}, S., {et~al.} 2007, \aap, submitted

\bibitem[{{Levshakov} {et~al.}(2005){Levshakov}, {Centuri{\'o}n}, {Molaro}, \&
  {D'Odorico}}]{levshakov05}
{Levshakov}, S.~A., {Centuri{\'o}n}, M., {Molaro}, P., \& {D'Odorico}, S. 2005,
  \aap, 434, 827

\bibitem[{{Lovis} {et~al.}(2006){Lovis}, {Mayor}, {Pepe}, {Alibert}, {Benz},
  {Bouchy}, {Correia}, {Laskar}, {Mordasini}, {Queloz}, {Santos}, {Udry},
  {Bertaux}, \& {Sivan}}]{lovis06}
{Lovis}, C., {Mayor}, M., {Pepe}, F., {et~al.} 2006, \nat, 441, 305

\bibitem[{{Mayor} {et~al.}(2003){Mayor}, {Pepe}, {Queloz}, {Bouchy},
  {Rupprecht}, {Lo Curto}, {Avila}, {Benz}, {Bertaux}, {Bonfils}, {dall},
  {Dekker}, {Delabre}, {Eckert}, {Fleury}, {Gilliotte}, {Gojak}, {Guzman},
  {Kohler}, {Lizon}, {Longinotti}, {Lovis}, {Megevand}, {Pasquini}, {Reyes},
  {Sivan}, {Sosnowska}, {Soto}, {Udry}, {van Kesteren}, {Weber}, \&
  {Weilenmann}}]{mayor03}
{Mayor}, M., {Pepe}, F., {Queloz}, D., {et~al.} 2003, The Messenger, 114, 20

\bibitem[{{Murphy} {et~al.}(2007{\natexlab{a}}){Murphy}, {Tzanavaris}, {Webb},
  \& {Lovis}}]{murphy07a}
{Murphy}, M.~T., {Tzanavaris}, P., {Webb}, J.~K., \& {Lovis}, C.
  2007{\natexlab{a}}, \mnras, submitted

\bibitem[{{Murphy} {et~al.}(2007{\natexlab{b}}){Murphy}, {Udem}, {Holzwarth},
  {Sizmann}, {Pasquini}, {Araujo-Hauck}, {Dekker}, {Fischer}, {Lovis},
  {Manescau}, \& {Pepe}}]{murphy07b}
{Murphy}, M.~T., {Udem}, T., {Holzwarth}, R., {et~al.} 2007{\natexlab{b}},
  \mnras, submitted

\bibitem[{{Murphy} {et~al.}(2003){Murphy}, {Webb}, \& {Flambaum}}]{murphy03}
{Murphy}, M.~T., {Webb}, J.~K., \& {Flambaum}, V.~V. 2003, \mnras, 345, 609

\bibitem[{{Norl{\'e}n}(1973)}]{norlen73}
{Norl{\'e}n}, G. 1973, \physscr, 8, 249

\bibitem[{{Palmer} \& {Engleman}(1983)}]{palmer83}
{Palmer}, B.~A. \& {Engleman}, R. 1983, {Atlas of the Thorium spectrum} (Los
  Alamos National Laboratory Report, LA-9615)

\bibitem[{{Pasquini} {et~al.}(2006){Pasquini}, {Cristiani}, {Dekker},
  {Haehnelt}, {Molaro}, {Pepe}, {Avila}, {Delabre}, {D'Odorico}, {Liske},
  {Shaver}, {Bonifacio}, {Borgani}, {D'Odorico}, {Vanzella}, {Bouchy},
  {Dessauges}, {Lovis}, {Mayor}, {Queloz}, {Udry}, {Murphy}, {Viel}, {Grazian},
  {Levshakov}, {Moscardini}, {Wiklind}, \& {Zucker}}]{pasquini06}
{Pasquini}, L., {Cristiani}, S., {Dekker}, H., {et~al.} 2006, in IAU Symposium
  232, ed. P.~{Whitelock}, M.~{Dennefeld}, \& B.~{Leibundgut}, 193--197

\bibitem[{{Whaling} {et~al.}(1995){Whaling}, {Anderson}, {Carle}, {Brault}, \&
  {Zarem}}]{whaling95}
{Whaling}, W., {Anderson}, W.~H.~C., {Carle}, M.~T., {Brault}, J.~W., \&
  {Zarem}, H.~A. 1995, Journal of Quantitative Spectroscopy and Radiative
  Transfer, 53, 1

\bibitem[{{Whaling} {et~al.}(2002){Whaling}, {Anderson}, {Carle}, {Brault}, \&
  {Zarem}}]{whaling02}
{Whaling}, W., {Anderson}, W.~H.~C., {Carle}, M.~T., {Brault}, J.~W., \&
  {Zarem}, H.~A. 2002, Journal of Research of the National Institute of
  Standards and Technology, 107, 149

\end{thebibliography}

\end{document}